%% file: main.tex
\documentclass[conference]{IEEEtran}
\usepackage{cite}
\usepackage{graphicx, subfigure}

\usepackage{algorithm}
\usepackage{amsmath}
\usepackage{algorithmic}
\usepackage{mathtools}
\usepackage{url}
\usepackage{listings}
\usepackage{xcolor}

\begin{document}
\title{APMEC: An Automated Provisioning Framework for Multi-access Edge Computing \\ \vspace{5mm} \Large Technical Report}

\author{\IEEEauthorblockN{
		Tung V. Doan\IEEEauthorrefmark{1},	
		Giang T. Nguyen\IEEEauthorrefmark{1} \IEEEauthorrefmark{2},
		Alexander Kropp\IEEEauthorrefmark{1}, and
		Frank H. P. Fitzek\IEEEauthorrefmark{1}}
	\IEEEauthorblockA{\IEEEauthorrefmark{1}Deutsche Telekom Chair of Communication Networks}
	\IEEEauthorblockA{\IEEEauthorrefmark{2}SFB-912 HAEC, Technische Universit\"at Dresden, Germany \\
		Email: \{firstname.lastname\}@tu-dresden.de}
}

\maketitle

\input{sections/abstract}
\input{sections/introduction}
\input{sections/related-work}

\input{sections/framework}
\input{sections/placement-algo}
\input{sections/realization}

\input{sections/evaluation}
\input{sections/conclusion}

\section*{Acknowledgment}
This work was supported in part by the German Research Foundation (DFG) within the Collaborative Research Center SFB 912 (HAEC) and BMBF-Project FAST.

\bibliographystyle{IEEEtran}
\bibliography{bib/references}

\end{document}

%% file: sections/abstract.tex
\begin{abstract}
Novel use cases and verticals such as connected cars and human-robot cooperation in the areas of 5G and Tactile Internet can significantly benefit from the flexibility and reduced latency provided by Network Function Virtualization (NFV) and Multi-Access Edge Computing (MEC). Existing frameworks managing and orchestrating MEC and NFV are either tightly coupled or completely separated. The former design is inflexible and increases the complexity of one framework. Whereas, the latter leads to inefficient use of computation resources because information are not shared. We introduce APMEC, a dedicated framework for MEC while enabling the collaboration with the management and orchestration (MANO) frameworks for NFV. The new design allows to reuse allocated network services, thus maximizing resource utilization. Measurement results have shown that APMEC can allocate up to 60\% more number of network services. Being developed on top of OpenStack, APMEC is an open source project, available for collaboration and facilitating further research activities.
\end{abstract}

%% file: sections/introduction.tex
\section{Introduction}
\label{sec:intro}

%
%
%


%
Tactile Internet~\cite{Fettweis-2014}, promisingly supports ultra-responsive and ultra-reliable applications such as steering and control of vehicles and industrial automation \cite{Simsek-2016} by providing extremely low latency.
However, latency reduction is limited by physical constraints such as geographical distance between clients and application server normally hosted at a faraway cloud platform. 
%
If application servers can be allocated in a close proximity to end users, at the mobile network edges, the latency can be significantly reduced. This solution implies to equip base stations with cloud computing capability to significantly increase its computation power to host applications.
That is the basic idea behind mobile edge computing, and later evolved into other access technologies, such as WiFi, becoming Multi-access Edge Computing (MEC). 
%
%
%
%
The challenges for MEC lies not only on hardware deployment at network edges and managing the virtualized resource, but also on software solutions to automatically orchestrate MEC applications and services and to coordinate with NFV frameworks responsible for virtualized network functions at the network core, which has been gathered significant momentum in research communities.

Even though management and orchestration (MANO) frameworks for NFV, such as Tacker~\cite{tacker} and OpenBaton~\cite{carella2015open} has been developed and tested, a counterpart for MEC has still been at horizon. Additionally, to maximize the interoperability to support users' mobility, such MEC framework has to support heterogeneous computation platforms with open and clearly defined interfaces to enable a variety of NFV frameworks.
Significant efforts have been done to define a clear and open architecture for MEC frameworks \cite{etsi-mec-framework}. The functional separation between manager, orchestrator as well as virtualization infrastructure manager makes it easer to design modular software solutions. 

However, a flexible design considering the interaction between MEC and MANO frameworks and a practical implementation and careful evaluation of such framework are still missing. The challenge is, therefore, complex and significant, especially because the development workload of such a framework is enormous.
%
%
%
%

In this paper, we design APMEC, a flexible and independent framework for MEC applications, yet allowing for the collaboration with NFV frameworks. 
To proactively orchestrate the resource allocation as well as life cycle related operations on both frameworks, we introduce the concept of a MEC service as the combination of a MEC application and its respective network service (NS).
This allows APMEC to bind new MEC applications to allocated yet underloaded network services to avoid creating additional network services, thus increasing computation resource utilization. 
Subsequently, APMEC includes a global orchestration module, namely MEC Service Orchestrator (MESO), which provides a common API to manage the MEC services. 
The design of our developed framework APMEC follows closely the MEC reference architecture of ETSI, with clearly defined interfaces to MANOs.

The contributions of this paper are threefold: \emph{First}, APMEC is a dedicated and flexible framework for MEC applications. APMEC's modular design follows closely ETSI's reference architecture for MEC, facilitating future extensions and supporting interoperability with other frameworks. Furthermore, APMEC's clearly-defined API to MANO modules support multi-site and multi-VIMs (Virtualization Infrastructure Managers).
\emph{Second}, the framework is ready to collaborate with existing and future MANO frameworks, such as Tacker, via a global orchestration module, increasing the number of allocated network services by 60\%, thus gaining resource utilization.
%
\emph{Third}, being developed on OpenStack, a carrier-grade VIMs and one of the most used OCCI \cite{occi} implementations, APMEC can potentially maximize its potential to be further developed or tested in real-world environment as well as open for research activities.

The rest of the paper is organized as follows. 
We discuss the background and related work in Section \ref{sec:related}, while Section \ref{sec:design} elaborates the architecture of the APMEC framework.
Then, we present the optimization problem of resource allocation and our heuristic algorithm in Section \ref{sec:optimization}.
Afterwards Section \ref{sec:evaluation} describes experiment setups and discusses evaluation results.
Finally, we concludes the paper and sketch our future work in Section \ref{sec:conclusion}.

%% file: sections/related-work.tex
\section{Background and Related work}
\label{sec:related}

In order to run applications at the network edge, Multi-access Edge Computing (MEC) needs to equip with cloud computing capability for application developers and content providers.
%
Thus, MEC needs to manage and orchestrate at the edge hosts all resources such as compute, storage and networking not only at host-level but also at system-level in an automated manner. The latter serves a central role and consists of four main building blocks as defined by ETSI's reference architecture \cite{etsi-mec-framework}.
%
%
%
The \emph{Virtualization Infrastructure Manager (VIM)} is responsible for managing, allocating and releasing virtualized resources to run software images of the MEC applications. 
The \emph{Mobile Edge Platform Manager} supervises the life cycle of applications, rules and service authentication.
\emph{Mobile Edge Orchestrator} maintains an oversight of the whole MEC systems, including topology, hosts' resources and assigns MEC applications to appropriate mobile edge hosts satisfying constraints on latency, available resources and services.
The fourth building block, \emph{Operations Support System (OSS)}, is the entry point receiving requests for application instantiation and termination, checks the validity of requests and forwards granted ones to the orchestrator.
%
%
%
Nevertheless, the realization of such framework to orchestrate and manage MEC applications, especially in coordination with network functions is still an open question.
\subsection{Requirements for a MEC provisioning framework}
\label{sec:requirements}

To automatically and efficiently manage and orchestrate MEC applications alongside existing MANO systems, the framework has to meet the following objectives: 

\begin{itemize}
	
	\item[R1]{\em Automation:} To automatically process the deployment requests, from translating them to deciding placement strategy, monitoring and deploying virtualized instances for MEC applications and network services.
	
	\item[R2] {\em Maximizing resource utilization:} efficiently use the overall computation resources by collaboration with MANOs.
	
	\item[R3] {\em Interoperability:} To allows the framework to work with a wide range of different VIMs.
	
	\item[R4] {\em Flexibility:} To operate either as a standalone entity or in collaboration with one or even multiple MANOs.
	
	\item[R5] {\em Small footprint:} To maintain a small and optimized code base by keeping only functionalities dedicated for MEC.
	
\end{itemize}

\subsection{Related work}
Considering the significant similarity between functional blocks of the MEC architecture and MANO for NFV~\cite{etsi-mano-framework}, a MEC framework should benefit from services offered by MANO. 
%
%
%
Several prior approaches have been proposed to manage and orchestrate MEC applications. 
Reusing MANO's manager and orchestrator for MEC applications is the approach proposed by Carella et al. \cite{Carella-2017}. The extension is mainly at the VIM entity to support container-based infrastructure. The small footprint of containers make them a viable candidate for deployment at edge nodes. 
The work is developed specifically on OpenBaton but similar extension can also be implemented for other MANO implementations such as Tacker \cite{tacker} and ONAP~\cite{onap}.
%
This approach, however, has several drawbacks: \emph{First}, OpenBaton has not been tested widely in large scales. \emph{Second}, the MEC framework is locked in a specific MANO implementation and less flexible for future MEC, thus hindering it from multi-MANO support. \emph{Third}, even though a prototype has been developed, no quantitative evaluation has been conducted.
In a different approach, NFV and MEC use separate managers and orchestrators, assuming that a common VIM is responsible for the virtualized infrastructure of both areas \cite{Sciancalepore-2016, li2017mec}.
MEC applications, therefore, can be flexibly provisioned either independently or in coordination with NFV. However, it is uncertain whether the architecture supports multiple MANOs. 
More importantly, the framework is not designed to efficiently provision resources in MANO framework.
Finally, there is neither prototype nor performance evaluation of the proposed architecture.
%

			
Orthogonal to the above approaches, a significant number of studies has been solving the challenge of service placement. Yang et al. \cite{yang2016seamless} addresses the challenge of MEC orchestrator and manager to optimally place MEC applications to physical hosts in the presence of a constantly changing workload.
Also tackling the placement optimization, Solozabal et al. \cite{solozabal2017design} chooses to modify the VIM entity, considering the mixed radio-cloud environment at the edge. 
Even though acknowledging the significance of those solutions, 
it is difficult to achieve optimal solution without considering both NFV and MEC.

All in all, is still an open research question w.r.t. an architecture and implementation to provision MEC applications in coordination with NFV in a resource-efficient manner. 
Furthermore, the framework has to support multi-VIM as well as multiple MANOs to cover the heterogeneity of the underlying networks. In the following we will describe APMEC, our automated provisioning framework for MEC applications addressing the above challenges.


%% file: sections/framework.tex
\section{APMEC: Framework Design}
\label{sec:design}

APMEC's hybrid approach is $i)$ to separate the orchestration and management between MEC applications and NS and $ii)$ to maintain a loose coupling with the MANO framework.
This allows for independently developing the two frameworks while keeping APMEC a small footprint by focusing only on MEC functionalities. At the same time, this enables APMEC to flexibly and efficiently decide connections between MEC applications and its respective network services.
%
In the following we elaborate our hybrid approach with ideas, concepts of APMEC and then its architectural design.

\subsection{Ideas and concepts}

\begin{figure}[t]
   \begin{center}
   \hspace*{-0.6cm}\includegraphics[trim = 8mm 60mm 8mm 75mm, clip, width=9cm]{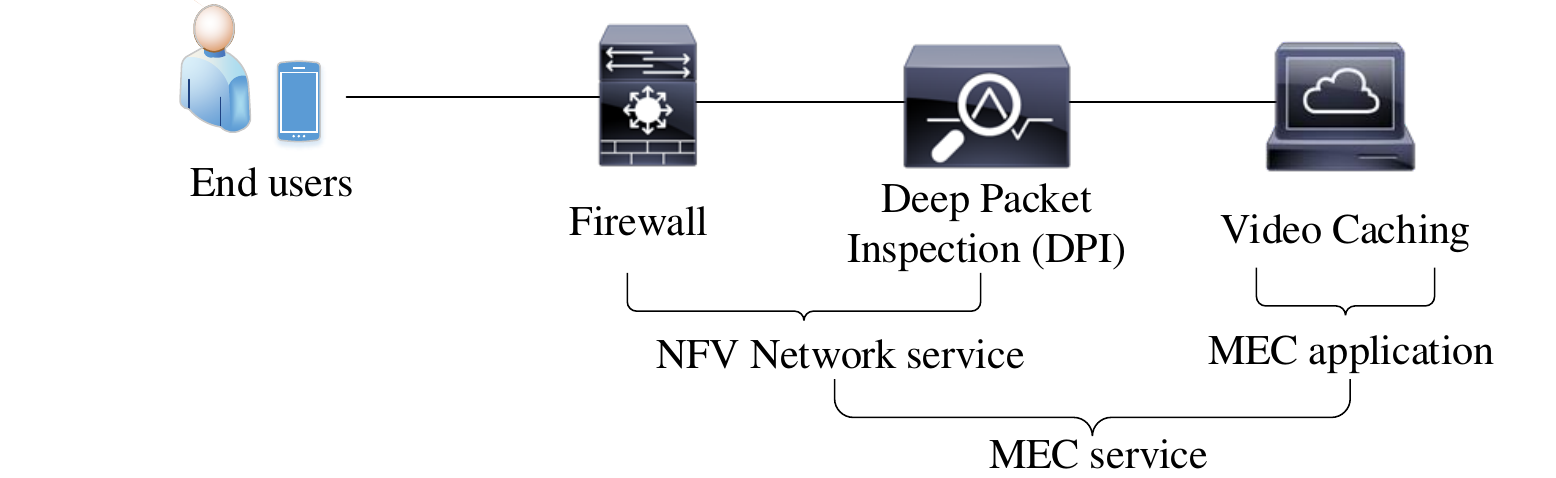}
   \caption{The concept of a MEC serice (MES), consisting of a Network service (NS) and a MEA (MEA) with an example of (Firewall-DPI) and Video Caching.}
   \label{fig:scenario}
   \end{center}
\end{figure}



To loosely couple MEC applications and network services, we
introduce the concept of a \emph{MEC service (MES)} which is a combination of $i$) a MEC application (MEA) and $ii$) a service function chain (SFC) consisting of a set of VNFs, represented by their identifiers. 
Fig.~\ref{fig:scenario} illustrates an example of a MES consisting of a video cache as MEC application and a NS which includes a Firewall and a Deep Packet Inspection (DPI). The NS lies between users and the MEC application.
%

To support the automated management and orchestration of MES, we introduce a unified data model to formulate resource requirements and orchestration scenarios. Data objects of the same format is then used for communication  between users and APMEC as well as internally between APMEC's components. 
%
%
This eliminates manual tasks and data conversion, thus reducing errors, overhead and latency.
Additionally, APMEC advocates an open and clearly design of its Application Programming Interface (API) to ensure interoperability between management and orchestration modules of APMEC and MANO frameworks. 
Subsequently, different MANOs can potentially communicate with APMEC via its API. 

\subsection{Architectural Design}

%

In accordance with the bottom-up approach when designing APMEC, we describe first its management and orchestration modules of MEC applications. The design of those modules follows closely ETSI's reference architecture for MEC \cite{etsi-mec-framework}. Afterwards we elaborate the interaction between APMEC and MANO frameworks and finally the interaction with the virtual infrastructure manager (VIM). 

\begin{figure}[t]
	\begin{center}
		\includegraphics[scale=0.4]{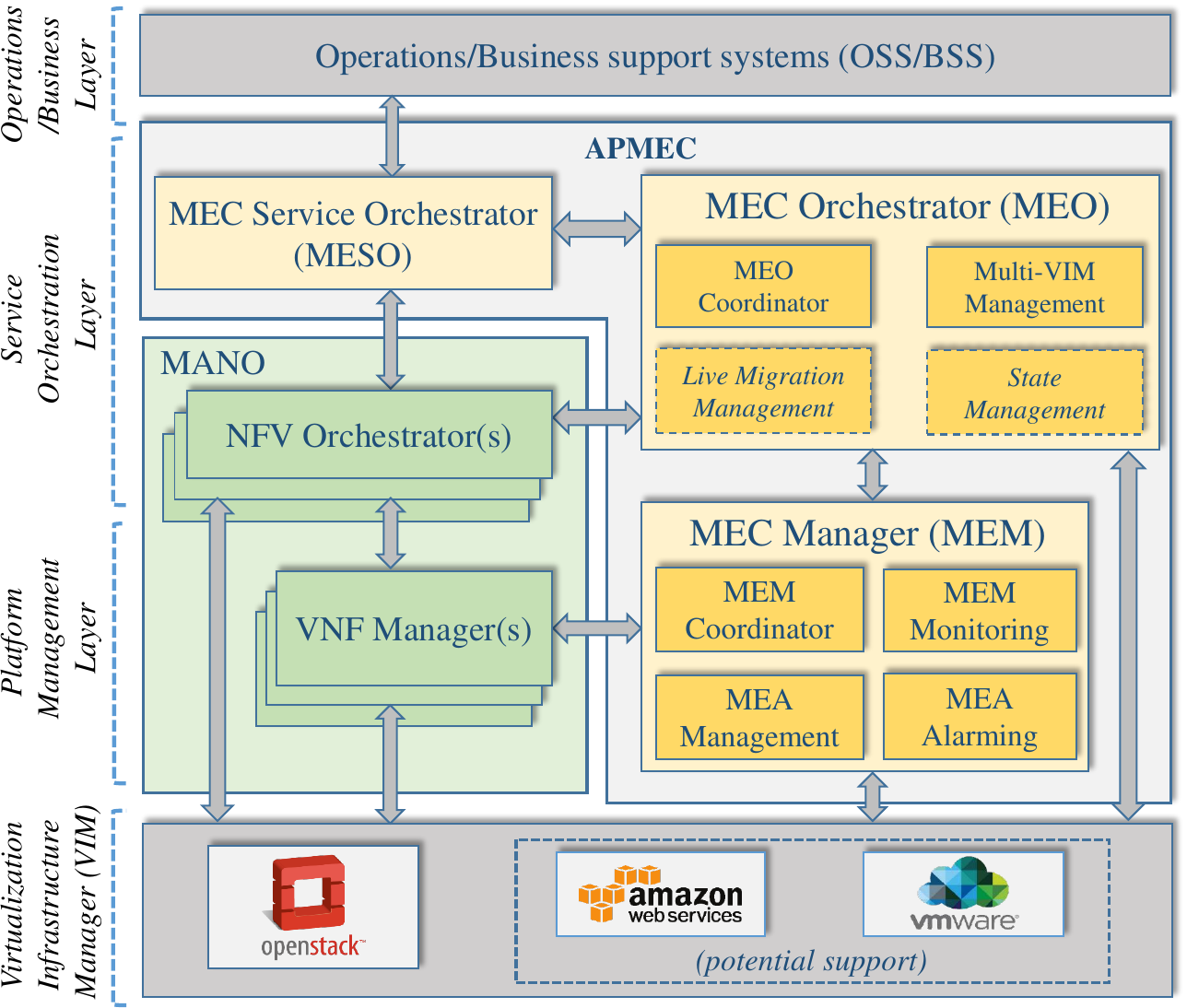}
		\caption{Architecture design of APMEC, in connection with Managment and Orchestration (MANO) and Virtual Infrastructure Manager (VIM).}
		\label{fig:framework}
	\end{center}
\end{figure}

\subsubsection{Management of MEC applications} 
{\em MEC Manager (MEM)} is in charge of MEA instances' life cycle, which includes automated provisioning, monitoring, configuration, healing, and scaling. They are managed by several modules as follows.
The automated provisioning functionality is in charge of receiving and dispatching users' requests for MEC applications. MEM provides an API for APMEC users to send their requests on MEC application's resources (CPU, memory, network, etc.). Then, the module translates those requests into a VIM's understandable format. Finally, MEM converts requests into VIM's function calls to actually instantiate MEA instances. 

\paragraph*{MEA Management} is in charge of configuring MEAs after they are launched. This normally involves the setting of parameters' values for bootstrap process.
%
%
For instance, a video caching server needs to know addresses of the original content server and other caching instances for collaboration.
%
%
%
%
%
In addition to that, MEA Manager is also in charge of automated scaling and healing. 
When MEA tends to be overloaded due to e.g. a surge in requests, MEA manager duplicates MEA instances using the stored information from the data objects.
Similarly, MEA manager relaunches MEA instances when the running ones fail.

\paragraph*{MEM Monitoring and MEM Alarming} To automate the scaling and healing process, MEM monitoring and MEM alarming components are introduced. 
%
%
%
APMEC receives from its users monitoring parameters and alarm configurations i.e. metrics, threshold, actions, etc. and the type of notification for each alarm.
%
When an alarm signal is trigger, due to e.g. a certain metric approaches its predefined threshold, the MEA alarming module calls appropriate procedures to react to the event.
Monitoring and alarming functionalities directly benefit the automated scaling and healing ones. 

\paragraph*{MEM Coordinator} is in charge of the direct cooperation between a MEA and VNF instances to improve efficiency. 
MEM Coordinator can request MANO to update the configurations of a specific network function.
E.g., via MEM coordinator video caching can directly request firewall--managed by VNFM--to add or update access rule or network policies. 
%








\subsubsection{Orchestration of MEC applications}
{\em MEC Orchestrator (MEO):} is mainly responsible for coordinating resources across various VIMs and deciding where to deploy MEAs. 
Through the API provided by MEM, MEO can make requests for the life cycle of the MEAs (initiation, deletion, update, etc.). Since MEO has global view of resources across multiple VIMs, MEM can also make a request to obtain VIM access from MEO so that the life cycle of the MEAs can be performed on the specific infrastructure. Such a design helps MEO to avoid bottleneck when managing the life cycle of the MEAs accross multiple VIMs.

\paragraph*{Multi-Vim Management (MM)} is in charge of communicating with multiple VIMs to support MEO's operation. This module should answer MEO's request on a list of VIMs satisfying a certain criteria, e.g. about distance or resource.
%
%

%

\paragraph*{MEO Coordinator} is responsible for coordinating the operations between NFV and the MEA. MEA can directly make an urgent request to update the NFV NS so that it helps improve the performance of the MEA immediately.
In particular, through APIs provided by the NFV Orchestrator, MEO Coordinator can directly make a request for NFV functionalities.
A particular MEA can directly make an urgent request to update the NS policy to quickly response to an incidence.
%
For example, the MEO coordinator can help MEC applications i.e. caching servers to request MANO's orchestrator to update its configuration to divert clients request through DPI when they observe a surge in request traffic.
%

\paragraph*{Live Migration Management (LMM) and State Management (SM)} are essential future modules for the live migration of running MEAs. To ensure the uninterrupted operation of MEA during migration process, their states require proper management. 

\subsubsection{Service Orchestrator (MESO)} is responsible for jointly coordinating the operations of MEO's and MANO's orchestrators. 
This module becomes the unique entry points to APMEC for user requests on resource of both MEC application and NS.
Without MESO, users need to interact with two frameworks for provisioning a service.
MESO's single interface simplifies further the operations of APMEC's users.
%
%
MESO provides dispatching functionalities to forward separately the requirements for APMEC and MANO modules. 
Since MESO has a broader view on both MEC and NFV, it is responsible for smart decisions on the placement of MEC applications as well as NSs to optimize for a set of requirements. 
The use of MESO also make it simpler for APMEC to interact with different MANOs. For each additional MANO framework, MESO might need to develop an adapted interface, while keeping the internal operations unchanged.
Using a unified data model, MESO allows users to describe the MES whose description contains information for both the requested NS and MEA.

\subsubsection{Other functionalities} 
An important functionality is the interaction of APMEC with Virtual Infrastructure Managers (VIMs).
In our design, APMEC and MANO frameworks can, but not necessarily, share the same infrastructure to leverage hardware resources.
Additionally, APMEC might also need to communicate with OSS/BSS module to coordinate its operations with legacy systems.
%
%
%

To summarize, this section presents APMEC framework that includes MEM and MEO to respectively manage and orchestrate the MEC applications and MESO to coordinate both MEA and NS.
To efficiently allocate computation resources for MEC services, MESO has to decide an optimized placement strategy whose algorithm will be detailed in the next section.

%% file: sections/placement-algo.tex
\section{Placement optimization of MEC services}
\label{sec:optimization}

 \begin{figure}[t]
    \begin{center}
   \includegraphics[trim = 70mm 65mm 50mm 65mm, clip, width=10cm]{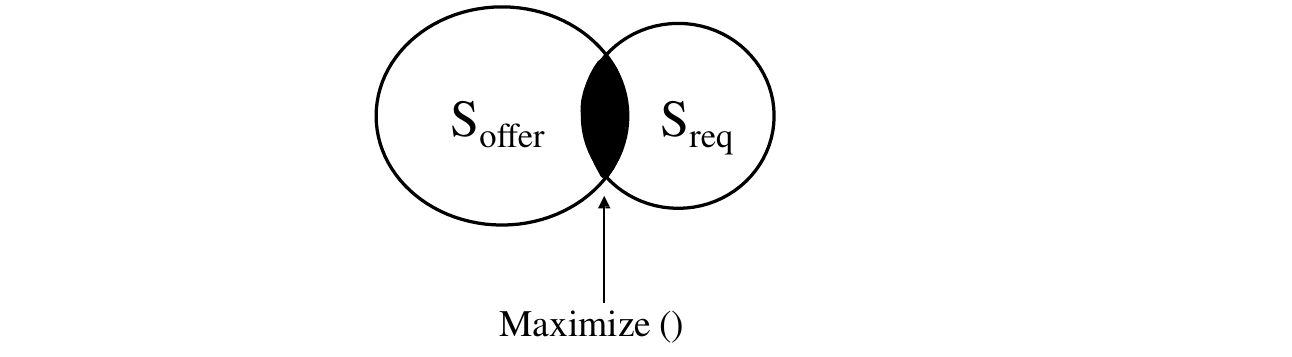}
   \caption{MEC service placement optimization problem which aims to maximize the overlapping area between the offered NSs $S_\text{offer}$ and the requested NSs $S_\text{req}$. }
   \label{fig:scenario}
   \end{center}
\end{figure}

Due to the fact that an NS is the combination of multiple NFs, the provisioning of the NSs could lead to the high deployment cost. It is worth noting that the MES includes the NSs and MEAs, therefore we solve the MES placement problem by reusing the NSs for multiple MEC applications. In this section, we first formulate the MES placement problem in terms of deployment cost optimization. We then propose a heuristic algorithm to minimize the number of virtual machines that are used for the MES initiation.

Let assume that $NF=\{NF_1, NF_2,...,NF_N\}$ is the set of $N$ network functions. For each $i \in \{1, 2,..., N\}$, $NF_i$ denotes the $i^{th}$ network function and ${y_i}$ denotes the number of instances of $NF_i$. Let $S_\text{offer}=\{s_1, s_2,...,s_M\}$ be the set of $M$ NSs that are offered by the MANO framework. For each $m \in \{1, 2,..., M\}$, $s_m$ denotes the $m^{th}$ NS.



In order to initiate the offered NS $s_m$, the number of NF instances could be formulated as:

\begin{equation}
F_{m}={\sum_{i=1}^{N}{x_{mi}}{y_i}}   \quad \forall m \in [1,M]
\label{eq:num-instances-in-ns}
\end{equation}\\
where:
\begin{equation}
x_{mi}=\left\{
\begin{array}{ll}
1 \qquad \mbox{if} \quad NF_i \in s_m,\\
0 \qquad \mbox{if} \quad NF_i \notin s_m.
\end{array} \right.
\label{eq:ns-condition}
\end{equation}

Let $K$ be the number of NSs $S_\text{req}=\{s_1, s_2,...,s_K\}$ that are requested to support $K$ MEAs $A=\{a_1, a_2,...,a_K\}$, repsectively. For each $k \in \{1,2,.., K\}$, $s_k$ and $a_k$ denote the $k^{th}$ requested NS and the $k^{th}$ MEA, respectively. The number of NF instances requested by NS $s_k$ is calculated as:

\begin{equation}
F_{k}={\sum_{i=1}^{N}{x_{ki}}{y_i}} \quad \forall k \in [1,K]
\label{eq:ns-offer}
\end{equation}\\
where:
\begin{equation}
x_{ki}=\left\{
\begin{array}{ll}
1 \qquad \mbox{if} \quad NF_i \in s_k,\\
0 \qquad \mbox{if} \quad NF_i \notin s_k.
\end{array} \right.
\label{eq:ns-condition}
\end{equation}

Let assume $F_{km}$ is the number of NF instances in the offered NS $s_k$ that are also re-used by the requested NS $s_k$. For each $k \in [1,K]$, $m \in [1,M]$, and $i \in [1,N]$, $y_{kmi}$ represents the number of NF instances of NF $i$ in the offered NS $s_m$ that could be re-used by the requested NS $s_k$. $F_{km}$ could be formulated as:

\begin{equation}
F_{km}={\sum_{i=1}^{N}{x_{kmi}}{y_{kmi}}}  \quad \forall k \in [1,K], \forall m \in [1,M]
\label{eq:ns-request}
\end{equation}\\
where:
\begin{equation}
x_{kmi}=\left\{
\begin{array}{ll}
1 \qquad \mbox{if} \quad NF_i \in s_k \cap s_m,\\
0 \qquad \mbox{if} \quad NF_i \notin s_k \cap s_m.
\end{array} \right.
\label{eq:ns-condition}
\end{equation}

From Eq.~\ref{eq:ns-offer} and Eq.~\ref{eq:ns-request}, without any placement optimization the total number of NF instances $T_{max}$ in the system could be given by:

\begin{equation}
T_\text{max} = \sum_{m=1}^{M}F_{m} + \sum_{k=1}^{K}F_{k}
\label{eq:total-instances}
\end{equation}

In this paper, we solve the placement optimization problem by minimizing the total number of NF instances $T$ in the system. This problem could be expressed as:

\begin{equation}
\large T \leq T_\text{max}
\label{eq:origin-problem}
\end{equation}

In order to solve the problem (\ref{eq:origin-problem}), we consider to reuse the offered NSs. Thus, the total number of NF instances $T$ could be formulated as:

\begin{equation}
T = \sum_{m=1}^{M}F_{m} + \sum_{k=1}^{K}F_{k} - \sum_{k=1}^{K}\sum_{m=1}^{M}F_{km}
\label{eq:total-instances}
\end{equation}

For each $i \in [1,N]$, let assume that $c_i$ is the reuse capacity of $NF_i$. Consequently, the problem (\ref{eq:origin-problem}) could be traformed to:
\begin{alignat}{5}
& \underset{k,m}{\text{max}}
& & \quad \sum_{k=1}^{K}\sum_{m=1}^{M}F_{km} \label{eq:np-hard}\\
& \text{s.t.} & &  \quad x_{ki} \in \{0,1\}, \forall k \in K, \forall i \in N, \label{eq:cons1}\\
& & & \quad x_{kmi} \in \{0,1\}, \forall k \in K, \forall m \in M, \forall i \in N, \label{eq:cons2}\\
& & & \quad c_i \leq C_\text{max}, \forall i \in N. \label{eq:cons3}
\end{alignat}\\
where $C_{max}$ is the maximum reuse capacity that each NF can obtain.




We acknowledge that the problem (\ref{eq:np-hard}) is a non-convex optimization problem since constraint (\ref{eq:cons1}) and constraint (\ref{eq:cons2}) are binary variables. Therefore, we proposed a heuristic algorithm to solve the aforementioned problem. We detail the proposed algorithm as follows.


To find the NS candidate, there are two conditions that need to be satisfied: $i$) the number of NFs in the NS candidate $s_m$ must be closest to the number of NFs in the requested NS $s_k$, and $ii$) the number of NF instances in the NS candidate $s_m$ have best fit to the number of NF instances in the requested NS $s_k$.

In order to comply with the first condition, the NS candidate $s_{m}$ could be used to serve the requested NS $s_k$, given by: 


\begin{equation}
m = \arg_m \max\sum_{i=1}^{N}{x_{kmi}} \quad \forall k \in [1,K], \forall m \in [1,M]
\label{eq:condition1}
\end{equation}\\
where:
\begin{equation}
x_{kmi}=\left\{
\begin{array}{ll}
1 \qquad \mbox{if} \quad y_{mi} - y_{ki} \geq 0,\\
0 \qquad \mbox{if} \quad y_{mi} - y_{ki} < 0.
\end{array} \right.
\label{eq:ns-condition}
\end{equation}

Eq.~\ref{eq:condition1} implies that the set of NFs in the requested NS $s_k$ should be mostly listed in the selected NS $s_{m}$ under the constrain of the number of NF instances as shown in (\ref{eq:ns-condition}). The Eq.~\ref{eq:condition1} could result in a set of NS candidates.

In order to optimize the number of NF instances, the number of NF instances in the NS candidate $s_{m*}$ have best fit to the number of NF instances in the requested NS $s_k$, given by:

\begin{equation}
m* = \arg_m \max\frac{F_{k}}{F_{m}}  \quad \forall k \in [1,K], \forall m \in [1,M]
\label{eq:condition2}
\end{equation}


\renewcommand{\algorithmiccomment}[1]{// #1}
\begin{algorithm}
\caption{The propsed algorithm for the MES placement}
\label{algo:apta-algorithm}

\hspace*{\algorithmicindent} \textbf{Input: $N, M, K, C_\text{max}$} \\
 \hspace*{\algorithmicindent} \textbf{Output: $m*$}
\begin{algorithmic}[1]
\STATE $k \gets 1$ \\
\COMMENT Phase 1: Finding the NS candidate that has the best fit to the numer of NFs in the requested NS $s_k$.
\WHILE{$k \leq K$}
\STATE $m \gets 1$ \\
\WHILE {$m \leq M$}
\STATE Find $P$ NSs that are satisfied with Eq.~\ref{eq:condition1}.\\
\STATE $m \gets m+1$
\ENDWHILE

\COMMENT Phase 2: Finding the NS candidate that has the best fit to the numer of NF in the requested NS $s_k$.
\WHILE{$m \leq P$}
\STATE Find NS $m*$ according to Eq.~\ref{eq:condition2}\\
\STATE $m \gets m+1$
\ENDWHILE

\STATE $k \gets k+1$
\ENDWHILE

\end{algorithmic}
\end{algorithm}

%% file: sections/realization.tex
\section{Realization of APMEC}
\label{sec:realization}

In this section, we describe the realization of the APMEC framework, including the modeling of data objects for inter-module communication, the initiation process and the implementation of APMEC.

\subsection{Data modeling for MEC services}
To enable the automated management and orchestration of MEC services,
we introduce a unified data model to formulate resource requirements
and orchestration scenarios. Communication between users and APMEC as
well as internal communication between APMEC's components use the same
format for data objects.  The unified data model eliminates manual
tasks and data conversion, thus reducing errors, overhead, and
latency.  We adopt the Topology and Orchestration Specification for
Cloud Applications (TOSCA)~\cite{tosca, 8255792}, an OASIS standard
language specification written in YAML format.  TOSCA supports an
automated management and is used in well-known NFV MANO frameworks,
such as Tacker~\cite{tacker}, OSM~\cite{osm}, and
OpenBaton~\cite{openbaton}, to describe the MEC services and the MEC
applications with a set of requirements.

\noindent\hspace{0.068\linewidth}\begin{minipage}{\linewidth}
\begin{lstlisting}[caption={TOSCA template for MEC service description}, label={lst:tosca}, captionpos=b]
tosca_definitions_version: tosca_simple_profile_for_mec_1_0_0
description: MEC service description
imports:
  meads:
    mead_templates:
      - mead1
      - mead2
    mec_driver: Apmec
  nsds:
    nsd_templates:
      - nsd1
    nfv_driver: Tacker
topology_template:
  node_templates:
      MEA1:
        type: tosca.nodes.mec.MEA1
      MEA2:
        type: tosca.nodes.mec.MEA2
\end{lstlisting}
\end{minipage}

We give a simple example to illustrate the TOSCA template of an MEC
service shown in Listing~\ref{lst:tosca}. The TOSCA template of an MEC
service contains the description of two MEC applications ({\em mead1}
and {\em mead2}) and one NFV network service ({\em nsd1}). Suppose
that {\em mead1} and {\em mead2} have already been created by APMEC
while {\em nsd1} is created by OpenStack Tacker.  Line 1 specifies the
TOSCA profile used for the MEC service description, whereby {\em
  tosca\_simple\_profile\_for\_mec\_1\_0\_0} contains the TOSCA
definitions for MEC, facilitating the validation of the MEC service
description in this example. For instance, {\em
  tosca\_simple\_profile\_for\_mec\_1\_0\_0} strictly requires that
all TOSCA templates for MEC must include keywords, such as {\em
  imports} and {\em topology\_template}, as shown in
Listing~\ref{lst:tosca}.  We use the OpenStack Tosca-parser which
includes a set of libraries to validate the TOSCA template before
passing the request to APMEC. For this reason, we added the support of
{\em tosca\_simple\_profile\_for\_mec\_1\_0\_0} to the
Tosca-parser~\cite{mec-tosca}. The addition of {\em
  tosca\_simple\_profile\_for\_mec\_1\_0\_0} to the Tosca-parser can
potentially be utilized by other OpenStack-related MEC projects, such as
StarlingX~\cite{starlingx} and Akraino~\cite{akraino}.


Lines 3--12 describe the {\em imports} section containing {\em meads}
and {\em nsds} which specify the TOSCA templates used for the MEC
applications and the NFV network services, respectively.  Lines 4--8
indicate that two TOSCA templates {\em mead1} and {\em mead2} are
imported for the creation of the MEC applications {\em MEA1} and {\em
  MEA2} and the specified onboarding framework (APMEC). Similarly,
Lines 9--12 indicate that the TOSCA template {\em nsd1} is imported
for the creation of the NFV network service {\em NS1}, which is then
dispatched to OpenStack Tacker (Line 12) via its API.  Lines 13--18
describe the topology template and its node templates, whereby {\em
  MEA1} and {\em MEA2} are two node templates, which correspondingly
invoke {\em mead1} and {\em mead2} defined in the {\em imports}
section (Lines 5--7).

\subsection{The Initiation Procedure for MEC services}

\begin{figure}[t]
	\begin{center}
		\hspace*{0.3cm}\includegraphics[trim = 5mm 100mm 150mm 60mm, clip, width=10cm]{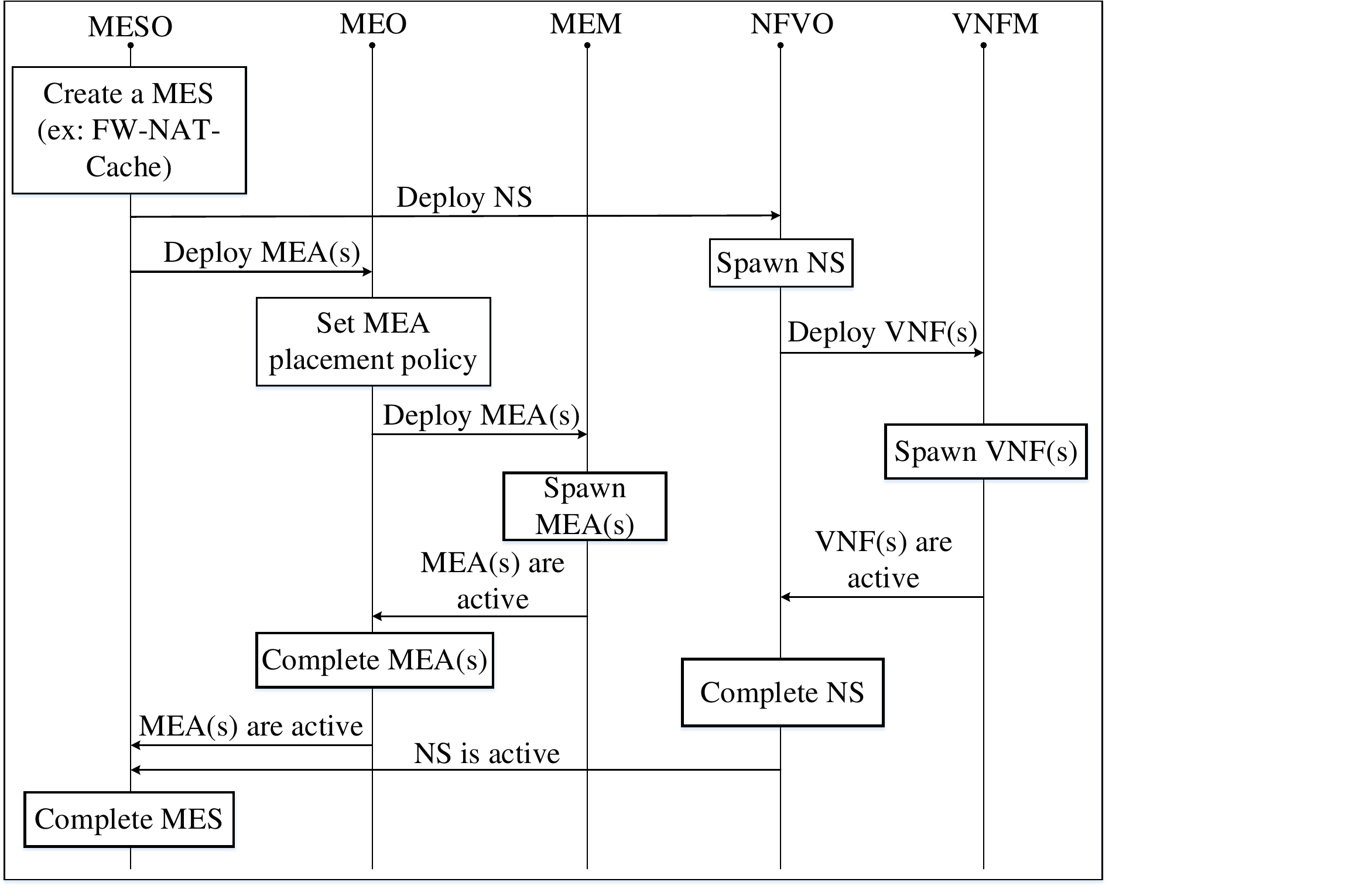}
		\caption{Initiation procedure for MEC services that shows the interoperability between APMEC and the MANO framework.}
		\label{fig:creation-procedure}
	\end{center}
\end{figure}

%
When APMEC receives via MESO a request for a MEC service, 
MESO parses the data model object in the request, break it into data segments for the MEA and network services. After running the placement algorithm, it performs the initiation procedure as follows.
First, MESO calls APMEC's MEO and MANO's NFVO to deploy MEA and network services respectively. 
For MEC applications, MEO sets the placement policy, which is the mapping of the virtual machines to physical hosts.  Afterwards, MEO requests MEM to instantiate the MEAs.
After validating the the validity of MEA description (i.e., by using openStack Tosca-parser to parse the MEA description), MEM calls the appropriate translator to convert the description from TOSCA format VIM's understandable format and then sends to VIM. Subsequently, VIM launches the MEA instances.
For network services, MESO requests NFVO to initiate the NS. NFVO, then, calls VNFM to instantiate the set of NFs included in the NS. 
The MES initiation is finished when both the MEAs and NS are successfully initiated.
The complete procedure of initiating MEC services is illustrated in Fig.~\ref{fig:creation-procedure}. 




\subsection{Implementation}



There are two options to implement APMEC, which is either starting from scratch or leveraging an existing framework. The former approach would provide more degree of freedom in the development process. However, an extensive effort has to be done to develop all the functionalities, including interfaces to various VIMs. APMEC follows the latter approach to leveraging the OpenStack framework \cite{openstack}.
OpenStack is the most used implementation of OCCI \cite{occi}, a specification aiming at smooth migrating cloud applications between cloud providers. Several operators have been planning to use OpenStack to manage resource their infrastructure to coordinate NS services \cite{operator1, operator2}. Building APMEC directly on top of OpenStack will increase the performance while leveraging the wide variety of supporting projects such as for monitoring and alarming among others.

%
In accordance with OpenStack's architecture, APMEC's code base is organized into three Python-based repositories: 
$i)$ Two {\em APMEC Clients} \cite{apmec-client, apmec-horizon} to support user interaction with the framework, either via command-line and web-based interfaces; and
$ii)$ An {\em APMEC Server} \cite{apmec-server} which is responsible for internal processing of requests, implementing functionalities detailed in the design section.
%
%
%
Management and orchestration modules such as MESO, MEO, and MEM are implemented as code modules of APMEC server. 
Each of those modules includes three elements: 
$i$) an API handler, which validates and dispatches user requests to the corresponding modules, 
$ii$) driver modules, which include libraries and APIs to perform specific functionalities (e.g., monitoring, alarming, etc.), and
$iii$) a plug-in module between the API handler and the driver, which processes user requests and calls appropriate drivers. 
This pluggable design allows developers to easily add new features to the framework (i.e., by adding the new drivers and connect them to the plug-in module).

%
The subsequent challenge is to translate this template into the formats understood by the underlying VIM managers. However, they mostly based on either JSON or YAML format and there exists tools to convert from one format to another.
%
%
%
Since APMEC is build on top of OpenStack, it has to translate TOSCA template into HOT template understood by {\em OpenStack Heat} \cite{heat}, the orchestrator for various services within OpenStack such as Nova for compute, Neutron for network, and Ceilometer for monitoring services etc.
%
Specifically, APMEC uses {\em Heat Translator} \cite{translator} for the translation between TOSCA and HOT templates. 
Future translators to support different VIMs, such as Amazon AWS or Microsoft Azure can be added with minor efforts. 
%

APMEC's monitoring functionality implements two services.
The fundamental one performs ping tests to confirm the reachability of a MEA instance. The advanced service allows for APMEC to collect a variety of metrics from a MEC application such as CPU workload and memory usage. APMEC leverages the available OpenStack Ceilometer \cite{ceilometer} projects for this functionality, implementing wrapper functions, facilitating the abstraction and extensions of new services.

To take advantage of the monitoring function, APMEC also includes alarming functionality that promptly informs APMEC about various events (e.g.,the MEAs get overloaded or halt). 
To enable the alarming functions, users first need to describe alarm configuration (e.g., CPU and memory threshold, metrics, etc.) for the specific MEA. Afterwards, APMEC generates an HTTP URI that contains the MEA identity corresponding to the alarm configuration through the alarm drivers. They are a set of libraries offered by the alarming tools. The alarm configuration and the corresponding URI are forwarded to Heat. In turn, Heat calls the API offered by the alarming tool to set alarm configuration. 
%
APMEC leverages OpenStack's Aodh project to provide the alarming services.

The auto scaling/healing of APMEC is implemented on top of the OpenStack Heat project. APMEC stores all the deployed Heat stacks defining resources for a certain application as well as its orchestration. 
They will be retrieved to recreate or duplicate instances up on request.
%


%
In our current implementation, APMEC -- especially MESO and MEO modules -- works closely with OpenStack Tacker as the backend manager and orchestrator for NS. However, APMEC is not bound to Tacker, but instead can potentially work with any MANO implementation given its API.
%
%



%% file: sections/evaluation.tex
\section{Performance Evaluation}
\label{sec:evaluation}

In this section, the developed APMEC framework and its best-fit heuristic algorithm for NS placement are assessed w.r.t its capability to efficiently utilize computation resource. 
Toward that goal we are interested in {\em the number of allocated network services} given a specific system capacity in terms of the number of virtual machines.

To verify the functional operation of APMEC, we first deploy the framework on a physical server with 32-core Intel Xeon CPU at 2.4 GHz and 128 GB of RAM spread across two NUMA nodes. The large amount of memory allows for the concurrent deployment of multiple VMs which is necessary for the experiment. The use of x86 CPU architecture simplifies the deployments due to the available of driver and software support for that CPU architecture. On top of the Ubuntu 16.04 LTS operating system, we then deployed OpenStack (version Pico). Tacker--the MANO framework for OpenStack--and APMEC were deployed afterwards. Tacker was used to deploy network services while APMEC was used for to deploy MEC alone or in combination with network services. All VMs running network functions have identical configuration with 1 vcpu, 10 GB hard disk and 500 MB RAM. 

After verifying the functional operation of APMEC on the practical testbed, we reverted to simulation to study its performance in a larger scale. Subsequently, we developed a lightweight simulator in Python, implementing only the NS placement algorithms of APMEC.
%
%
Initially, the system is empty without any network service. The simulation starts with the first network service's request. Any subsequent request for a new network service is generated one after another when the previous request had been allocated. Each requested network service was generated with a random size. The process stops when the system has no capacity to allocate a new request.
%
We fixed the system capacity to 100 VMs. 
%

%
We ran the simulation to compare two approaches representing the interaction between MEC and NFV frameworks: $i)$ separated frameworks and $ii)$ cooperated frameworks. The two approaches are referred to as \emph{separation} approach and \emph{cooperation} approach hereafter.

\subsection{Impacts of the sizes of network services}

\begin{figure*}[t]
	\begin{center}
		\subfigure[The number of accepted requests versus NS size.] {
		\label{fig:ns_size}
		\includegraphics[width=8cm]{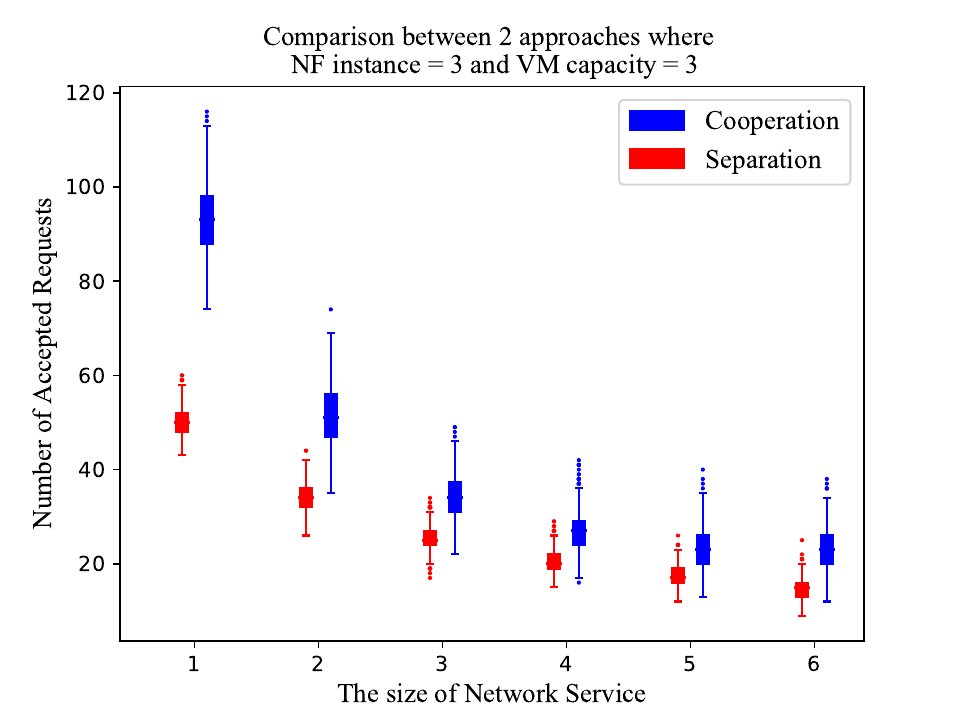}}
        \subfigure[The number of accepted requests versus NF instances.]{
        \label{fig:nf_instances}
        \includegraphics[width=8cm]{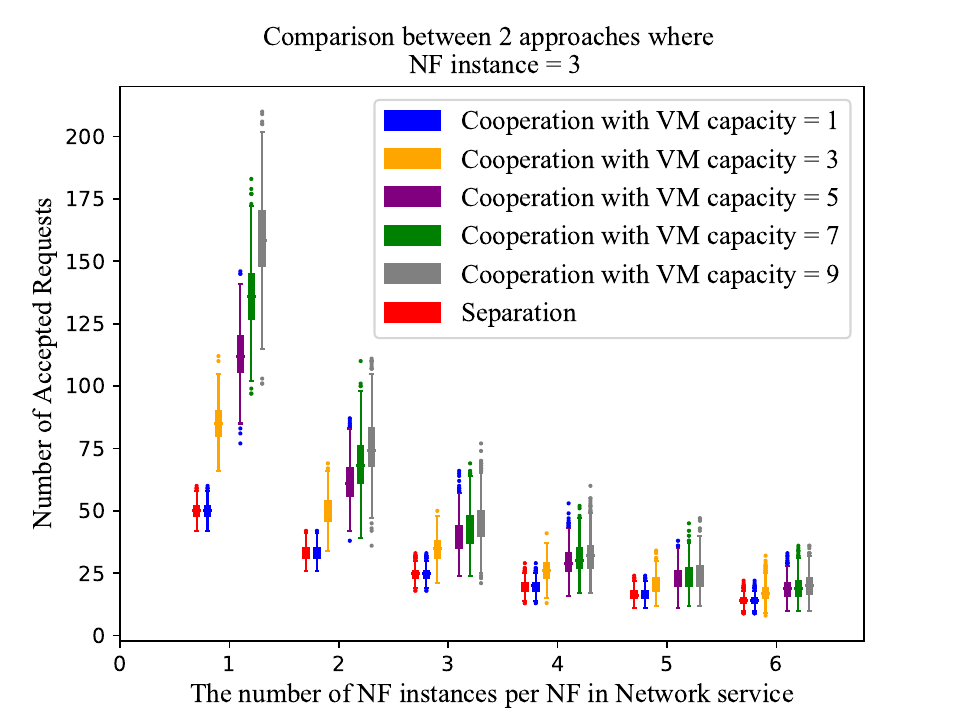}}
 		\caption{The number of accepted requests are used for the MES creation compared between {\em Cooperation} and {\em Separation} approaches.}
		\label{fig:requests}
	\end{center}
\end{figure*}

We would like to investigate the impact of the size of network services, meaning the number of contained network functions, to the two approaches. Subsequently, we fixed the number of NF instances and the VM capacity to three while varying the maximum size of requested network service between one and six. The number of allocated network services for both approaches were collected and plotted in Fig.~\ref{fig:requests}(a). 
%
As the maximum sizes of network services increases, the number of number of allocated network services for both approaches decreases. However, the magnitude of decrease slows down as the size increases.
In all settings, the cooperation approach outperforms the separation one.
In intuitively, the reason for this achievement is that the separation approach does not consider to reuse the NSs, therefore it rapidly consumes the hardware resources in terms of VMs. Subsequently, it results in the reduction of the number of allocated network services. 
In contrast, the cooperation approach leverages the running VMs of deployed newly arriving NSs, reserving more available VMs for later network services.


%


\subsection{Impacts of the NF instance and VM capacity}
In this part of the experiments, we would like to investigate the impact of NF instance and VM capacity to the two approaches. 
Subsequently, we fixed the maximum size of network service to three while varying the NF instance between one and six and the VM capacity between one and nine with a step of three.
The number of allocated network services for both approaches were collected and plotted in Fig.~\ref{fig:requests}(b).
%
As NF instance increases, the number of allocated network services for all systems' configuration decreases. 
However, in all settings, the cooperation approach outperforms the separation one. Additionally, as the VM capacity increase, the number of allocated network services in cooperation approach is significantly larger than that of the separation approach.
%
With a reasonable NS size of three and the VM capacity of five, meaning five network services can share one VM of a particular network function, the cooperation approach and allocate almost 60\% more number of network service as compare to the separation approach.

%% file: sections/conclusion.tex
\section{Conclusion}
\label{sec:conclusion}

In this paper, we design APMEC, a novel and dedicated framework for automated provisioning of MEC applications.
The design of APMEC allows for jointly collaborating with MANO frameworks.
Subsequently, we introduce the concept of a MEC service, consisting of a MEC application and a Network Service (NS), allowing for reusing allocated network services for new requests, thus maximizing computation resources. 
After modeling the network service placement as an optimization problem, the paper proposes a best-fit heuristic. APMEC and its heuristic are developed on top of OpenStack.
Experiment results have shown that APMEC can allocate up to 60\% more number of network services.

The framework paves the way for several future research directions, including: $i$) the implementation of more comprehensive optimization to further improve resource utilization and $ii$) the live migration of MEC applications.

%
%
%